# Facile synthesis of potassium intercalated p-terphenyl and signatures of a possible high $T_c$ phase

P. Neha, V. Sahu*, and S. Patnaik*

School of Physical Sciences

Jawaharlal Nehru University, New Delhi, India

**Abstract**

Synthesis methodology for flakes of p-terphenyl through sublimation under inert atmosphere of argon is presented. Flake morphology of p-terphenyl provides a favourable environment for efficient intercalation of potassium. Ratio of potassium and p-terphenyl is adjusted so as to obtain the desired superconducting phase i.e. potassium doped p-terphenyl ($K_3C_{18}H_{14}$). A clear transition is observed at 107 K under Zero Field Cooled (ZFC) and Field Cooled (FC) mode. But overall the moment is positive possibly due to impurity phase dominating characteristics in the presence of negligible superconducting volume fraction. The M-H loop taken at 20 K shows magnetic behaviour in synthesized K- doped p-terphenyl but upon background subtraction, it does exhibit characteristics of a type-2 superconductor.

Corresponding authors:

rbsvikrant@gmail.com, spatnaik@mail.jnu.ac.in

## Introduction

The search for high temperature superconductivity in organic macromolecules has received a strong impetus with the detection of diamagnetic transitions around 125 K in potassium doped para-terphenyl ($C_{18}H_{14}$) [1]. Predicted theoretically by W. A. Little (way back in 1964)[2], superconductivity in carbon based compounds were first detected in graphite intercalates around 1965 [3] and had achieved a maximum Transition temperature ($T_c$) around ~ 38 K in $Cs_3C_{60}$ . In the recent past, superconductivity in alkali metal doped aromatic carbon compounds (such as picene, coronene, phenanthrene, and 1,2.8,9 dibenzopentacene) have attracted significant attention. Intercalation of various metal atoms in solid-phase polycyclic aromatic hydrocarbon (PAHs) like coronene and picene have resulted in unconventional superconductors with $T_c$ above 15 K[4,5]. In unaltered solid form, layers of picene molecules are arranged in a herringbone manner into which metal ions can be staked between *ab*-layers. The resulting structure possesses quasi 2D structure having properties similar to inorganic superconductors like cuprates, iron-pnictides, $MgB_2$ etc.[6-8]. Furthermore, under the general gamut of resonance valence model, strong correlation physics has been invoked to explain superconductivity in doped PAH [9]. Emergence of superconductivity in these organic compounds have been characterized as a combination of Boson and Fermion components of Mott insulating phase with underlying charge transfer and consequent doping [9].

Recent revival of research in high $T_c$ organic compounds relates to its discovery in series of compounds from different groups with $T_c$ in the range of 5 K to 125 K for K-doped p-terphenyl ($K_3C_{18}H_{14}$) [10-12]. Para -terphenyl is a fairly common industrial compound that is known to be present in a derivative form in edible mushroom [13], in organic scintillators [14,15] and it is also used as a precursor for the synthesis of picene [5]. To be realistic, superconductivity claims in alkali metals doped PAHs have always been questionable due to their non-reproducible nature. Furthermore, the diamagnetic ring currents of Benzene and its derivatives are easy to get misunderstood as Meissner diamagnetism of stable superconductors, particularly when the superconducting volume fraction is low. Nevertheless, a sharp change in diamagnetic moment in a majority paramagnetic background phase cannot be assigned to such possibilities and this is what has been repeatedly observed in p-terphenyl [1]. Moreover, being extremely susceptible to degradation in air and moisture, the cause for inconsistency in results could be much more than variation in stoichiometry, homogeneity and phase purity of sample [16]. Thus the reaction temperature and ambient conditions play a vital role in synthesis because harsh temperature and pressure can alter the structure of PAHs [16-18]. Therefore, the main challenge in preparing the K-doped PAHs is to optimize the conditions for its synthesis so that doped specimen can lead to reproducible results. In this paper we focus on developing a facile route for synthesis of K doped p-terphenyl towards its possible confirmation as a high temperature superconductor. P-terphenyl flakes are grown in argon atmosphere through sublimation and slow cooling. Magnetic measurements on K-doped p-terphenyl flakes confirm a possible diamagnetic transition around 107 K.

## Results and discussion

The potassium intercalated p-terphenyl samples were synthesised by taking potassium metal (>99% purity, Loba Chemie ) and p-terphenyl (>99% purity, Alfa Aeser). The powder p-terphenyl was first purified through sublimation before doping with K. Synthesis of K doped p-terphenyl sample is a two step process. In the first step we tried to attain p-terphenyl crystals. The entire process was carried out in argon atmosphere inside a glove box to get rid of degrading agents such as oxygen and moisture. In glove box sublime P-terphenyl was taken in borosilicate petri-plate that was covered with a borosilicate lid. Then the covered petri-disc was kept on a hotplate and temperature of the hotplate was increased to slightly above the boiling of the P-Tephenyl (~150 C). On observing the onset of boiling, the heating was stopped and the petri-plate was allowed cool down to room temperature. The cover-lid acted like a substrate over which vapour deposited shiny white flakes of P-terphenyl could be seen. These P-terphenyl shiny crystalline flakes were recovered from borosilicate lid. P-terphenyl is extremely hygroscopic and extreme care needs to be taken to avoid exposition to air while synthesis and measurements. Schematic illustration of K doped p-terphenyl synthesis along with reaction chart is shown in Figure 1. Optical microscope image ( in Varian 7000 Raman) of p-terphenyl crystalline flake is shown in inset of Figure 1. All characterization of K doping synthesis are carried out with these crystalline flakes. Evidence for stacked layers similar to graphite is seen in these optical images. The bigger crystals had a size of ~100 micron and had clean and shiny surface texture implying imbedded metallic characteristics.

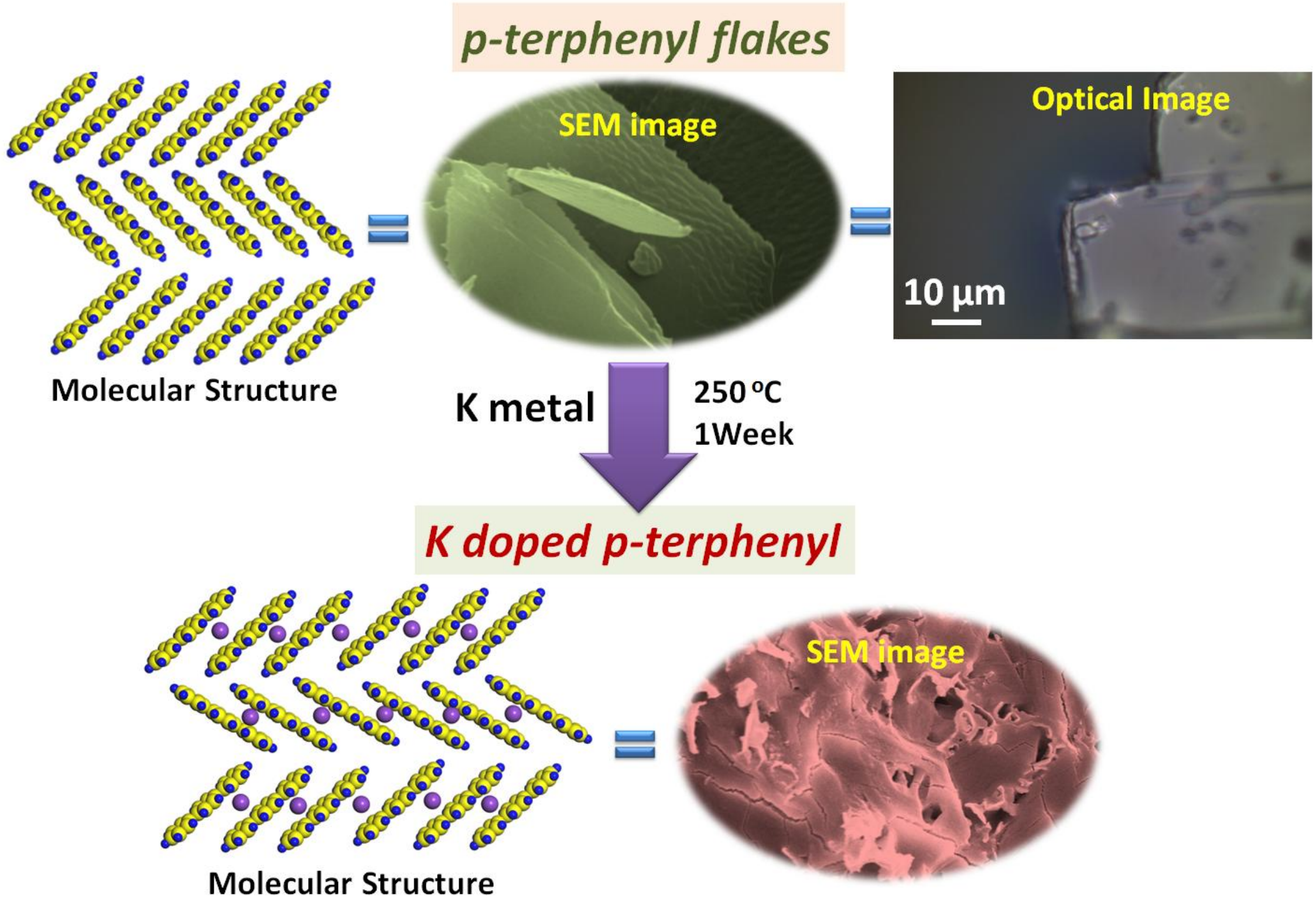

Figure 1. Schematic illustration of K doped p-terphenyl synthesis.

Next we discuss synthesis procedure for K-doped p-terphenyl (with possible superconducting phase). In the first step we used the sublime p-terphenyl flakes with measured amount of K metal pieces in a quartz tube and sealed it in vacuum (~ 1 mbar). This is a method commonly used for intercalation in other organic compounds such as picene and coronene and this method does not involve physical mixing. After heating at 250$^{o}$C for one week we could not achieve a homogeneous compound. Moreover, un-reacted p-terphenyl was clearly visible. The magnetic characterizations also ruled out any trace of superconducting signal. Evidently, potassium could not be intercalated onto p-terphenyl through this method. Subsequently, we tried the mixing method where-in we thoroughly mixed the K metal pieces with P-terphenyl with the help of mortar and pestle in argon atmosphere. Once the homogeneous mixture was formed, it was transferred to a quartz tube and vacuum sealed. It sealed quartz tube was then heated at 250$^{o}$C for 7 days and cooled down slowly over one day. The compound formed was grey in colour. The magnetic characterizations ruled out superconductivity even in this set of samples. It was concluded that potassium intercalation has remained marginal if at all. This guided us to make changes in the synthesis strategy. This time, we used only lustrous K metal pieces (by removing the whitish covering from the K metal which is generally formed when stored in inert liquid) for physical mixing into single crystalline flakes of p-terphenyl. After mixing with p-terphenyl flakes, in mortar-pestle it was transferred to quartz tube for sealing in vacuum. Sealed tube was kept at 250$^{o}$C for 7 days. Compound formed at this time was completely black in appearance and also showed signs of diamagnetic transition at elevated temperatures. Obtained K-doped p-terphenyl crystals were further characterized with structural characterization techniques that will be discussed later. But the main conclusion from our synthesis methodology is that even the traces of possible oxidized potassium at the surface of metal pieces can result in failure of attaining possible superconducting phase.

In Figure 2, FTIR spectra of pure p-terphenyl flakes recorded on BX Perkin Elmer instrument are shown.

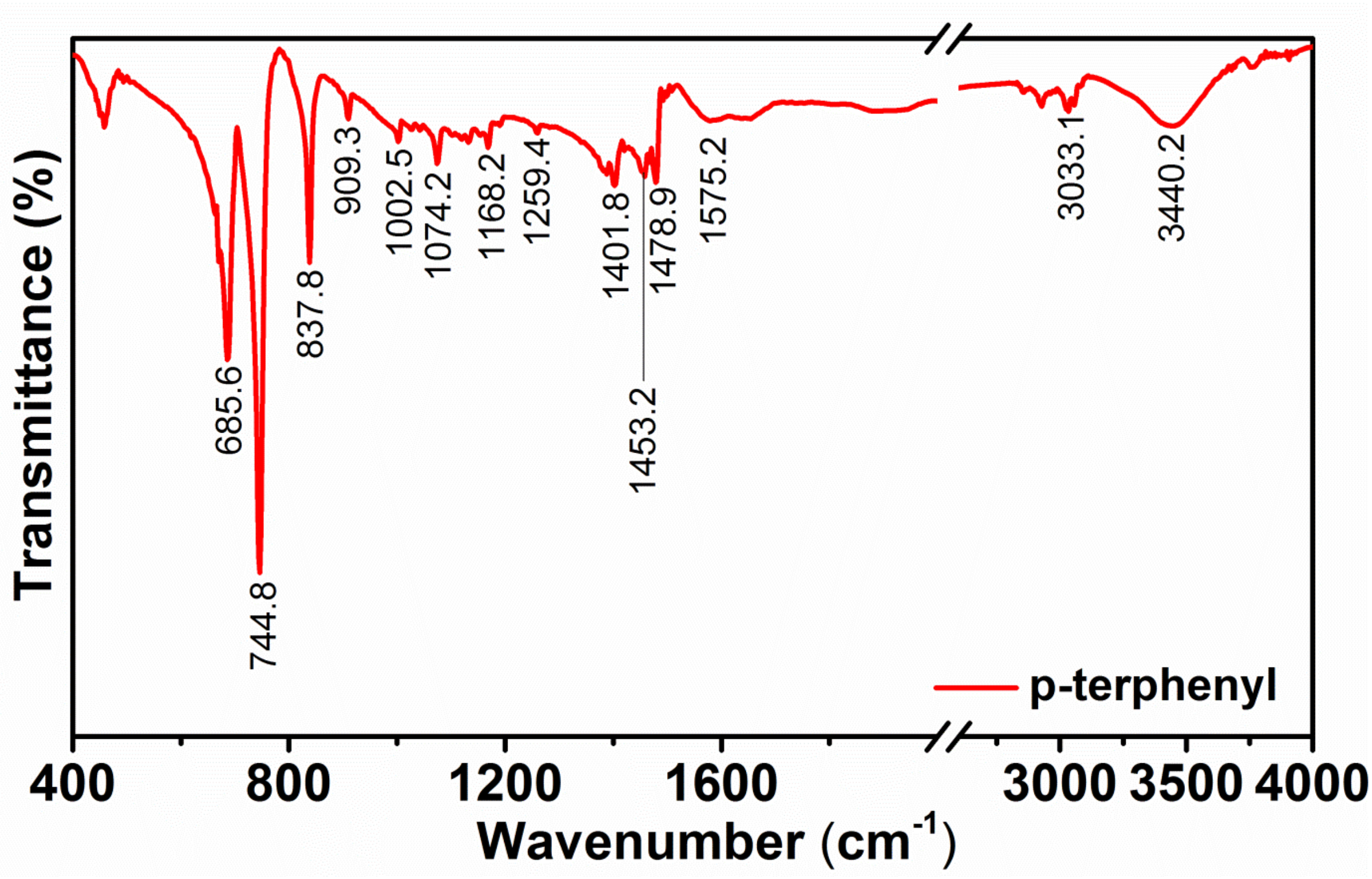

Figure 2. FTIR spectrum of p-terphenyl crystalline flakes.

Structural identification of p-terphenyl flakes was carried out using the FTIR spectroscopy (Figure 2). This optical transmission spectrum of p-terphenyl matches well with the single crystal FTIR spectrum of p-terphenyl reported earlier [19]. Prominent peaks at 658.6 and 744.8 $cm^{-1}$ belongs to terminal rings C-H bending vibrations while terminal ring C-H bending mode is reflected at 837.8 $cm^{-1}$. Transmittance peaks below 1300 $cm^{-1}$ correspond to the out-of-plane and in-plane wagging vibrations. Four sharp peaks at 1401.8, 1453.2, 1478.9 and 1575.2 $cm^{-1}$ arise due to the vibration of ring skeletal respectively. Overtone and combination peaks appear between 1666 and 2000 $cm^{-1}$. Moreover, C-H stretching of connected benzene rings gives impression of a broad band at 3033.1 $cm^{-1}$.

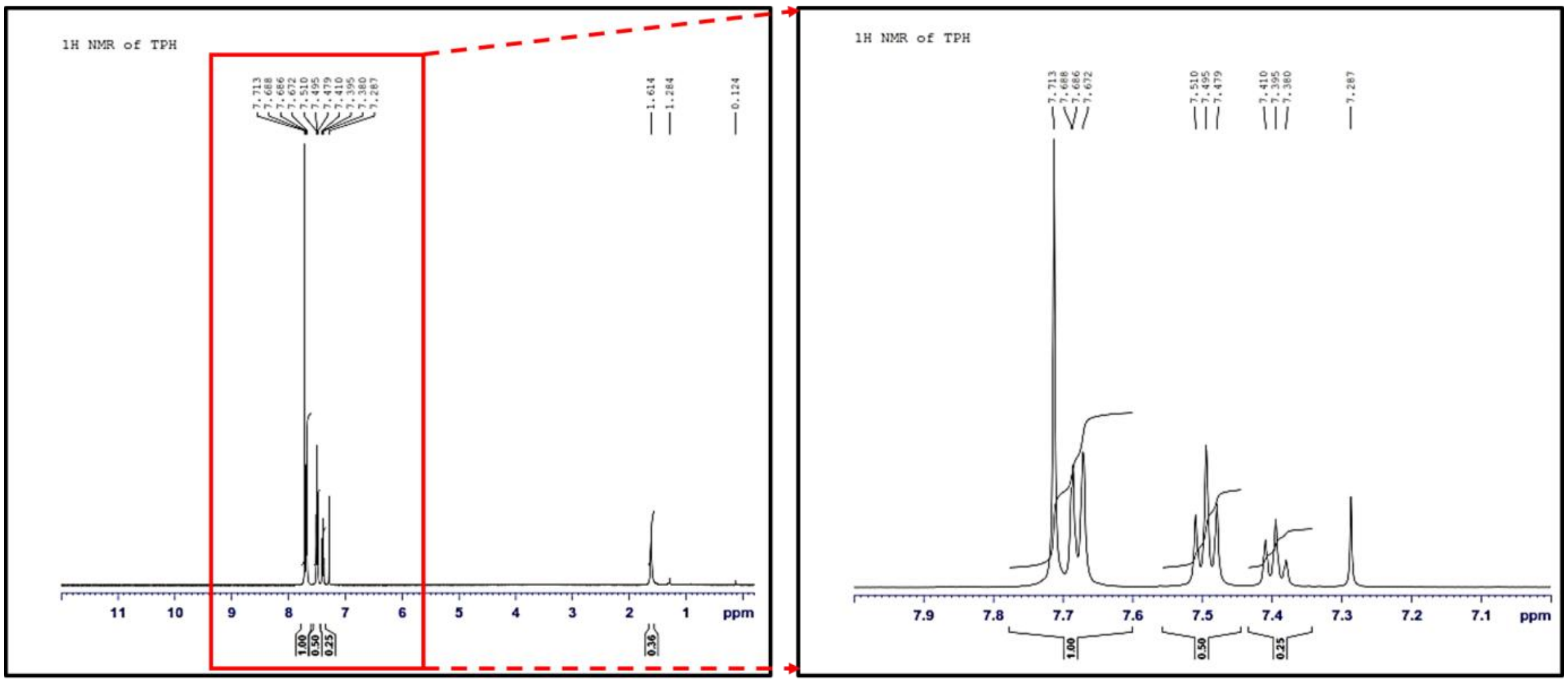

Figure 3. $^{1}$H NMR spectrum of p-terphenyl crystalline flakes.

Figure 3 shows the $^{1}$H NMR spectrum of as grown p-terphenyl crystalline flakes. Singlet sharp peak at 7.713 ppm corresponds to the four protons present on central aromatic ring. [19] 7.688 and 7.686 ppm doublet correspond to ortho-position proton of terminal rings. Triplet peak at 7.510, 7.495 and 7.479 ppm in spectrum arise due to the meta-position proton in terminal ring. While the next triplet at 7.410, 7.395 and 7.380 ppm are of para-position terminal proton. Presence of four different types of proton in p-terphenyl confirm from the integral ratio of the peaks (2:2:2:1) and also the spectrum result emphasize high purity of synthesized p-terphenyl sample.

The X-Ray diffraction pattern of pure p-terphenyl and K doped p-terphenyl are shown in Figure 4. Pristine specimen shows planes with well-defined peaks. Analysis confirms monoclinic crystal structure for P-terphenyl single crystal flakes [20]. Doped sample showed slightly amorphous character with diffused peaks. Disappearance of 001 peak in K doped specimen (within our low angle limitation) implies layer separation has increased, reflecting a possibility of successful intercalation of K between layers of p-terphenyl.

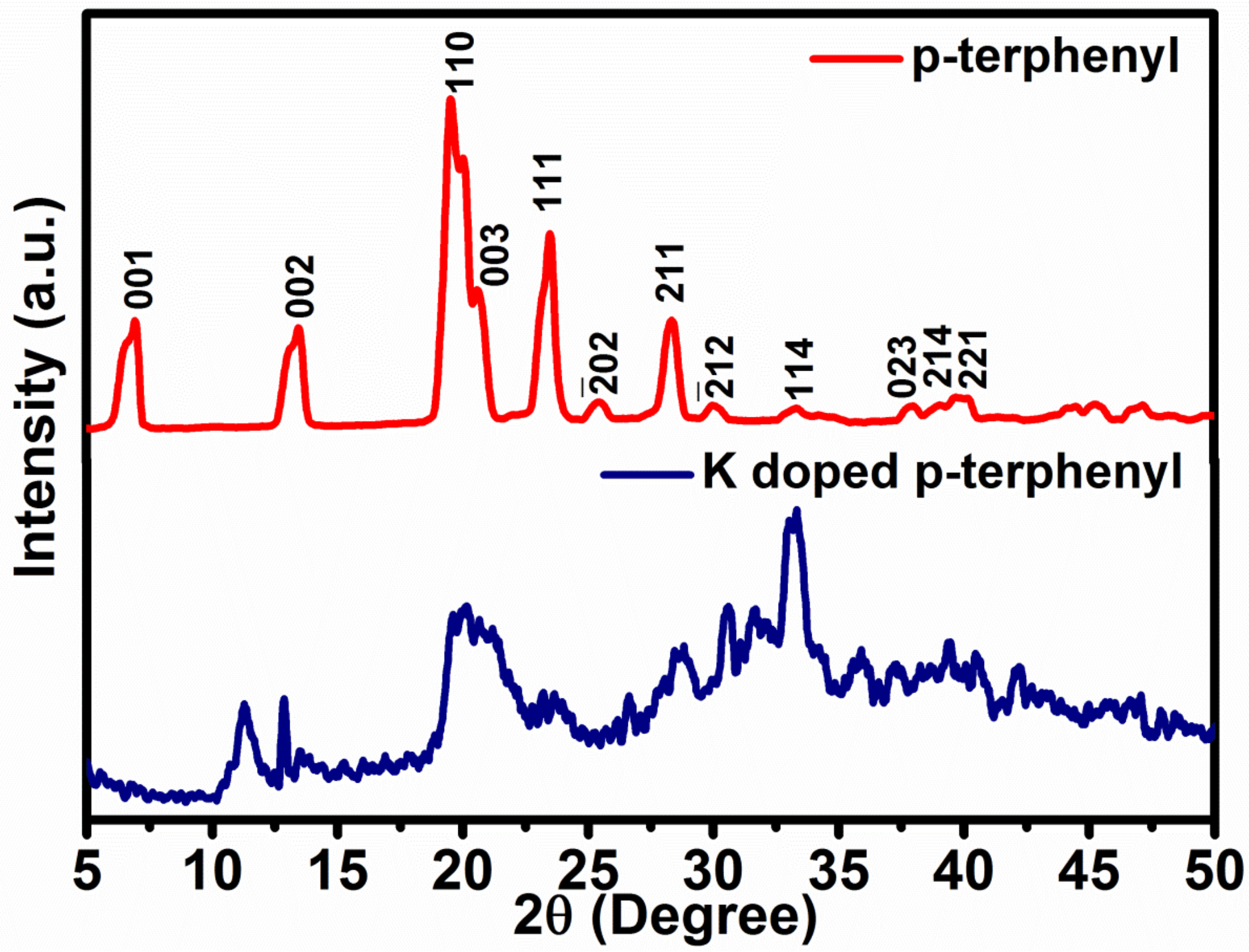

Figure 4. The XRD pattern of pristine sublimed p-terphenyl and K intercalated p-terphenyl are shown by red and blue lines respectively.

To consolidate structural characterization of as grown and doped samples, non-destructive Raman spectroscopy was carried out to study the vibrational modes of P-terphenyl and its changes with K doping (Figure 5). A sharp peak at high wave number(3061.9 $cm^{-1}$) is seen only in the pure P-terphenyl that arises due to the vibration of C-H. This is not seen in K-doped P-terphenyl. We note that these measurements were taken in air and while both pure and doped samples are highly hygroscopic, the potassium doped sample is more susceptible to degrade because of formation of metal oxides. Further, the bare sample of p-terphenyl shows ring skeletal vibration of C-C bond at 1597.3 $cm^{-1}$ and the counterpart K doped p-terphenyl shows slight upper shift because of doping (1599.3 $cm^{-1}$). Similar upper shift is observed in inter C-C stretching in pristine (1273.2 $cm^{-1}$) and doped (1275.3 $cm^{-1}$) p-terphenyl. Peaks in the lower shift region correspond to C-H bending (~800 -1200 $cm^{-1}$) and C-C-C bending (~400-800 $cm^{-1}$). From the relative graph it is evident that K doping in p-terphenyl affect the structure of bare material which result in low intensity peak of C-H bending and C-C-C bending [9].

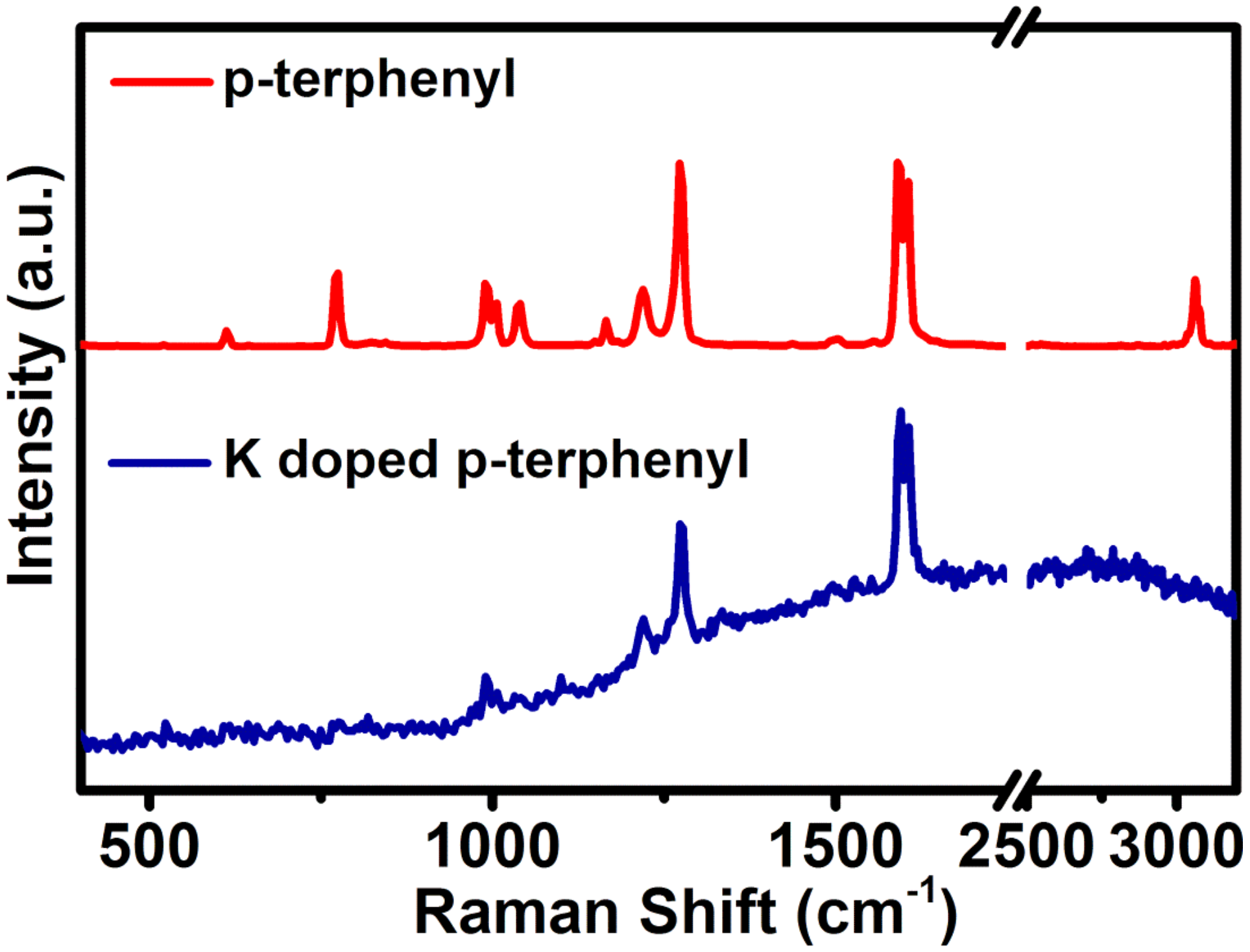

Figure 5. Raman spectra of sublimed p-terphenyl and K-intercalated p-terphenyl are shown in red and blue colour lines respectively.

To investigate the morphology variation of both un-doped (6a and 6b) and doped (6c and 6d) samples of p-terphenyl, SEM micrographs are compared in Figure 6. Micron scale p-terphenyl flakes is clearly visible in Figure 6a. Further at higher magnification strata of p-terphenyl are evidently noticeable (figure 6b). Reduction in the size of flakes resulting from the doping is markedly seen in figure 6(c and d) which is in agreement with similar observation reported in K-doped poly aromatic hydrocarbons [10]. Layered structure is somewhat less evident with K intercalation in the layers of p-terphenyl possibly due to underlying tendency to amorphousize the compound. Moreover, porosity was found be low in K-doped sample that is also reflective of the chemical changes occurring during intercalative process followed for K doped p-terphenyl.

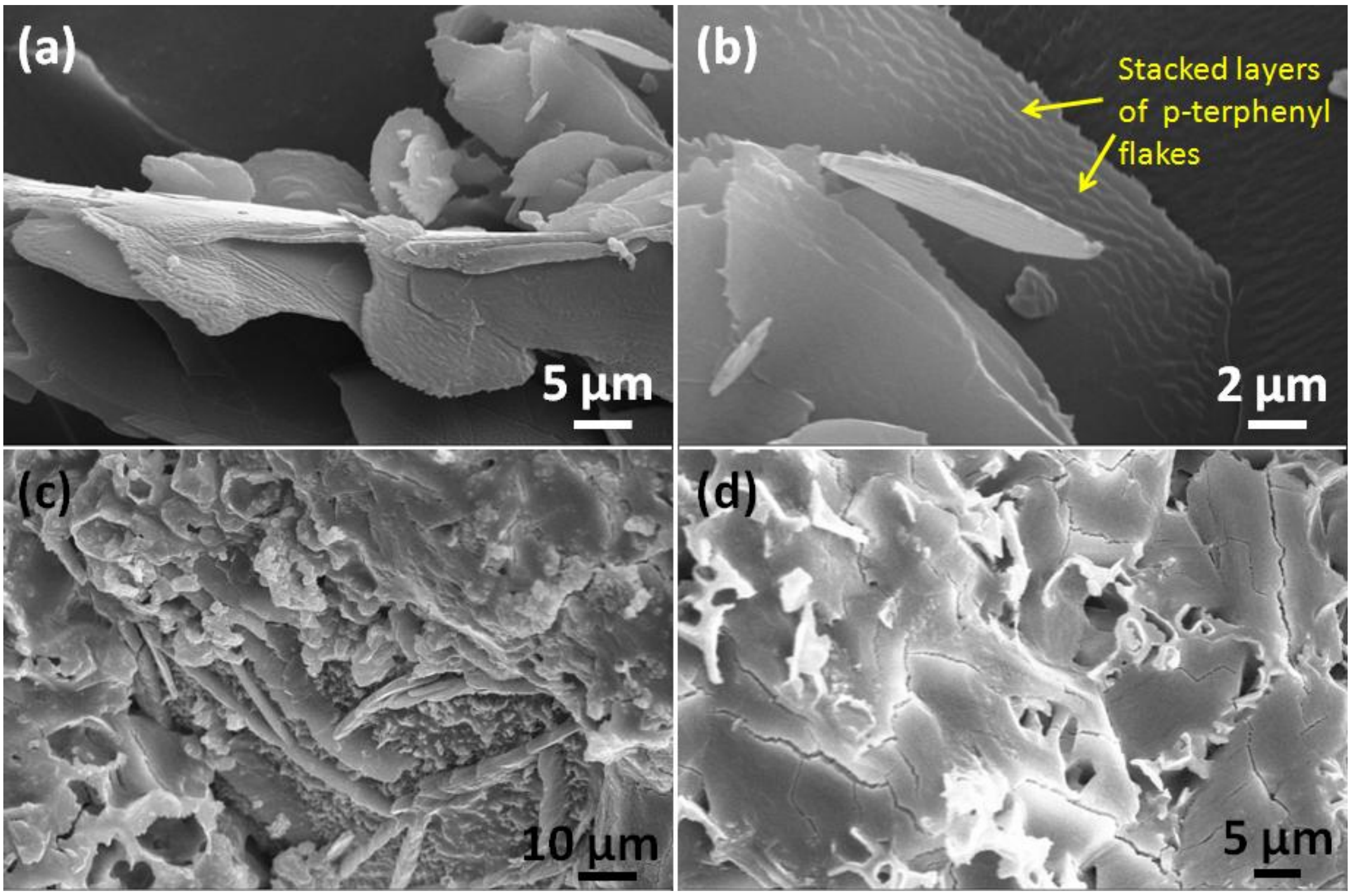

Figure 6. SEM images of sublimed p-terphenyl (a and b) and K-intercalated p-terphenyl (c and d).

The temperature dependent magnetic behaviour of synthesised K doped p-terphenyl material is illustrated in Figure 7. We followed the ZFC-FC protocol to decipher superconductivity which is described below. Initially the synthesised material was cooled down to low temperature (down to 2K) in zero field cool (ZFC) mode and then an external field of 50 G was applied. The sample was and then heated to 120 K at the rate of 1K/min (Fig 7a, Blue) in the presence of field and data (ZFC) recorded. Then the sample was cooled down to 2K under applied field of 50G and heated again to 120K with the field remaining unchanged (FC, Fig 7a Red). We note that both in ZFC and FC data a sharp change in moment is observed around 107 K. However the overall moment is positive which is in consonance with earlier reports where the sample was synthesized in under high pressure [1].

In lower temperature range, variation of magnetic moment with temperature is reflective of overall dominance of paramagnetic phase. It is evident that (if at all), these PAH based superconductors have very low superconducting volume fraction. The inset of figure 7(a) shows the variation of moment with temperature after cooling under ZFC mode at applied field of 50 G when pure P- terphenyl data (at 50 G) are subtracted from K-doped p-terphenyl data. The pure p-terphenyl is a paramagnet down to lowest temperature. On subtraction of background a clearer transition is observed at 107 K which partly becomes diamagnetic in absolute scale. In figure 7b the suppression of this transition on the application of higher magnetic field is studied. Data for two ZFC field value of 50 G and 500G are chosen and the magnetization axis is normalized with its value at 120 K. Evidently, on increasing the magnetic field the possible transition at 107 K is significantly suppressed. These scans are not back-ground corrected and no inference should be done on the upper critical field of these PAH superconductors. The main inference is that the transition at 107 K is relatively suppressed with a 10 fold increase in external field. Such variation in moment with increasing field is a characteristic feature for type-II superconductors. In p-terphenyl presence of superconductivity is also reported at 5K, 43 K and 120 K with K intercalation [10-12].

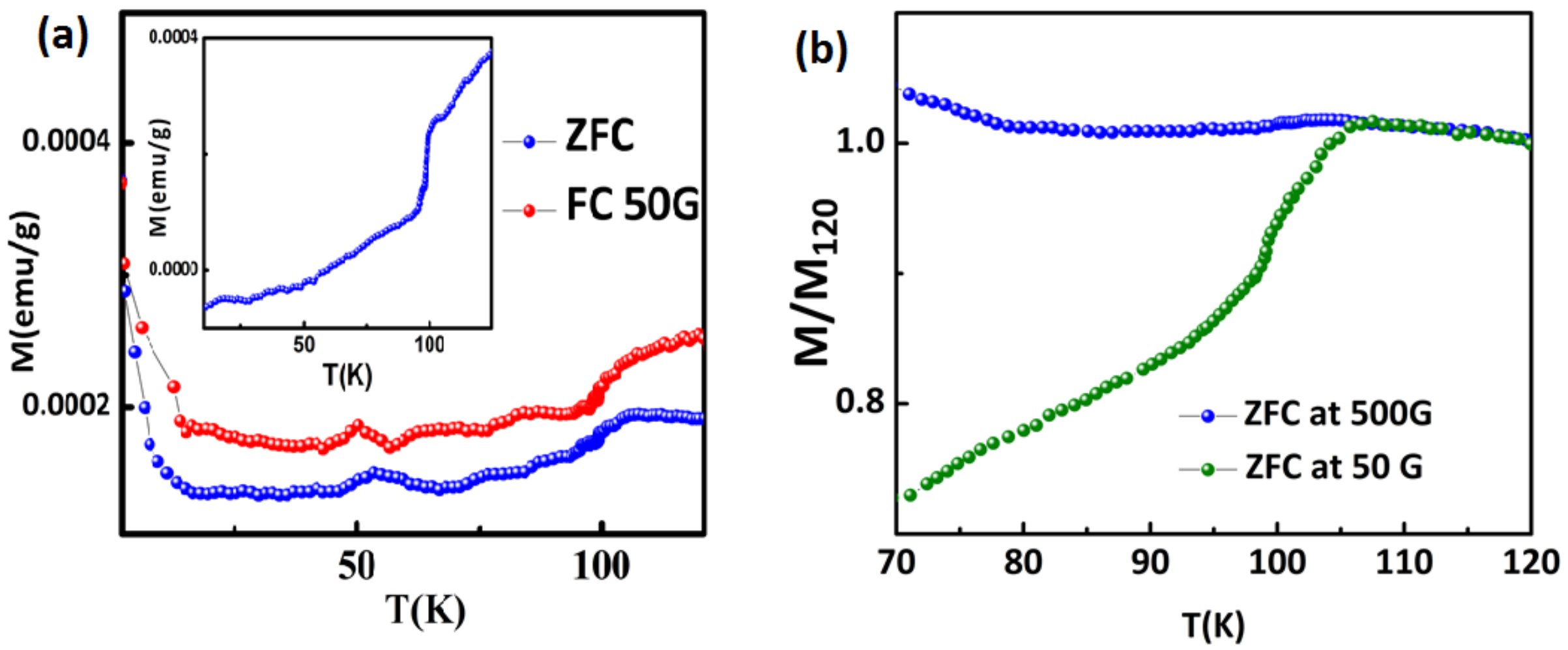

Figure 7. (a) Magnetization variation as a function of temperature. The inset shows the data after background subtraction.(b) Normalized M-T curve taken at 50G and 500G.

Figure 8(a) shows the variation of magnetic moment as a function of applied magnetic field at constant temperature of 20 K for K doped p-terphenyl. It indicates some sort of a overall ferromagnetic loop behaviour but its origin is yet to be ascertained [1]. The M-H loop curve doesnot saturate at higher field range like ferromagnetic materials, instead the moments are decreasing in higher field range which can be ascribed to diamagnetic component present in the synthesized compound. By taking the average of the loop value (corresponding to increasing and decreasing field values) as the back ground, in Figure 8b we plot the magnetization vs. external field (M-H loop) for K doped p-terphenyl and the results are strikingly similar to what is expected for a type- II superconductor. Even features of low field diamagnetic behaviour with upturn corresponding to lower critical field is confirmed.

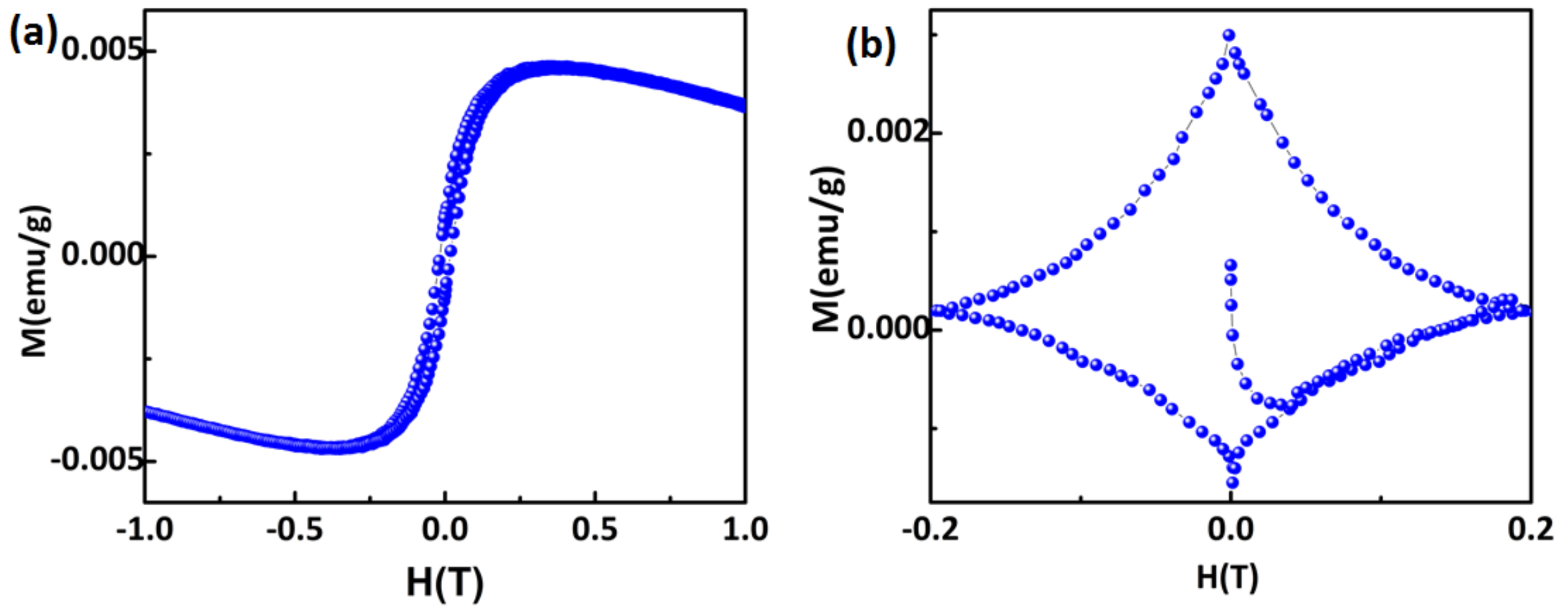

Figure 8. M-H loop of K doped p-terphenyl taken at 20 K. (b) M-H loop of K doped p-terphenyl taken at 20 K after background subtraction.

## Conclusion

In conclusion, we report synthesis of K doped p-terphenyl by a simple method. The magnetic measurements on as grown crystals consolidate the possibility of a lurking high temperature superconductivity phase at an elevated transition temperature of $T_c$ ~107 K. X-Ray Diffraction and Raman spectroscopy measurements performed on pristine and doped specimens relate to the effect of K doping in the compound. The magnetic measurements under ZFC-FC mode provide indication of presence of possible superconductivity in strong paramagnetic background and show evidence of some sort of transition around 107 K. The M-H loop taken at 20 K depicts the magnetic nature of the synthesised K-doped p-terphenyl but under background subtraction, a typical type-II superconductivity feature is revealed.

**Acknowledgement**:

We thank Dr. V P S Awana (from NPL, New Delhi India) for suggestions and discussions. Mr. Soroj Jha at AIRF (JNU) is specifically thanked for taking the high resolution magnetization data for all the samples.